\RequirePackage{lineno} 

\documentclass[twocolumn,preprintnumbers,amsmath,amssymb,nofootinbib]{revtex4}
\usepackage{graphicx}% Include figure files
\usepackage{dcolumn}% Align table columns on decimal point
\usepackage{bm}% bold math
 
\usepackage{epstopdf}
\def\bea{\begin{eqnarray}}
\def\eea{\end{eqnarray}}

\def\pp{\mbox{$p$-$p$}}

\def\pa{\mbox{$p$-A}}

\def\auau{\mbox{Au-Au}}

\def\ppb{\mbox{$p$-Pb}}

\def\pn{\mbox{$p$-N}}
\def\aa{\mbox{A-A}}
\def\nn{\mbox{N-N}}

\def\pt{$p_t$}

\def\yt{$y_t$}

\def\nch{$n_{ch}$}
\def\mmpt{$\bar p_t$}

\begin{document} 

\setlength{\pdfpagewidth}{8.5in}
\setlength{\pdfpageheight}{11in}

\setpagewiselinenumbers
\modulolinenumbers[5]

\addtolength{\footnotesep}{-10mm}\
% \linenumbers

\preprint{version 0.2\textsl{}}                                             

\title{Some physics of small collision systems
}

\author{Thomas A.\ Trainor}\affiliation{University of Washington, Seattle, WA 98195}

%%%%%%%%%%%%%%%%%%%%%%%%%%%%%%%

\date{\today}

\begin{abstract}
In recent years certain experimental results from small collision systems (e.g.\ p-p, d-Au, p-Pb) at the RHIC and LHC have been reinterpreted as evidence for formation therein of a dense flowing medium (QGP) despite small collision volumes. Systems that had been assigned as simple references (e.g.\ cold nuclear matter) for larger A-A collisions would then no longer play that role. This presentation examines conventional interpretations of certain data features in the context of a two-component (soft+hard) collision model. Specific topics include centrality determination for p-Pb collisions, interpretation (or not) of nuclear modification factors, significance of claims for strangeness enhancement, and interpretation of the ``ridge'' in p-p collisions. For p-p and p-Pb data analysis results indicate that p-Pb collisions are simple linear superpositions of p-N collisions, and N-N collisions within small systems generally follow simple and consistent rules. However, there is more to be learned about ``basic" QCD in small systems with improved analysis methods.
\end{abstract}

%\pacs{12.38.Qk, 13.87.Fh, 25.75.Ag, 25.75.Bh, 25.75.Ld, 25.75.Nq}
%\keywords{Suggested keywords}

\maketitle

%%%%%%%%%
\section{Introduction}

This presentation addresses recent claims that a quark-gluon plasma (QGP) forms in smaller collision systems (e.g.\ \pp\ and \pa) despite small space-time volumes~\cite{phenix,gardim}. Such claims are contrary to assumptions prior to RHIC operation that small systems would provide control experiments against which \aa\ results might be compared to test claims of QGP formation for the larger systems. In what follows a two-component (soft plus hard) model of hadron production in high-energy collisions provides the basis for examining phenomena related to QGP claims. Issues considered include (a) centrality determination for \pa\ collisions, (b) interpretation of so-called nuclear modification factors (NMFs), (c) strangeness enhancement as a signal for QGP formation and (d) the CMS ridge, a feature of 2D angular correlations, as indicating ``collectivity'' (flow) in \pp\ collisions.

%%%%%%%%%%%%%%
\section{TCM for (multi)strange spectra}  \label{spectrumtcm}

The two-component model (TCM) is  derived empirically from \pt\ spectrum evolution with  charge multiplicity \nch. The soft component represents projectile-nucleon dissociation along the beam axis. The hard component represents minimum-bias parton fragmentation to jets.

As a preamble to defining a spectrum TCM A-B collision centrality must be established such that event charge density $\bar \rho_0$ is decomposed into soft and hard components $\bar \rho_0 \equiv n_{ch}/\Delta \eta = \bar \rho_s + \bar \rho_h$ where for an A-B collision system $\bar \rho_s = (N_{part}/2) \bar \rho_{sNN}$ and $\bar \rho_h = \ N_{bin} \bar \rho_{hNN}$ including participant nucleon N pair number and \nn\ binary collision number. For \nn\ binary collisions empirically-inferred relation $\bar \rho_{hNN} \approx \alpha \bar \rho_{sNN}^2$ is of fundamental importance.

Given a \pt\ spectrum TCM for unidentified-hadron spectra based on centrality determination~\cite{ppprd,newpptcm} a corresponding TCM for identified hadrons may be generated by assuming that each hadron species $i$ comprises certain {\em fractions} of TCM soft and hard total particle densities $\bar \rho_{s}$ and $\bar \rho_{h}$ denoted by $z_{si}(n_s)$ and $z_{hi}(n_s)$ so that $\bar \rho_{si} = z_{si}(n_s) \bar \rho_{s}$ and $\bar \rho_{hi} =z_{hi}(n_s) \bar \rho_{h}$~\cite{ppbpid,pidpart1,pidpart2,pppid}.
The PID spectrum TCM is then 
\bea \label{pidspectcm}
\bar \rho_{0i}(p_t,n_s) &=& S_i(p_t) + H_i(p_t,n_s)
\\ \nonumber &\approx&  \bar \rho_{si} \hat S_{0i}(p_t) +   \bar \rho_{hi} \hat H_{0i}(p_t,n_s),
\eea
where such {\em factorization} is a fundamental aspect of the TCM.
Unit-normal model functions $\hat S_{0i}(m_t)$ (soft) and $\hat H_{0i}(y_t)$ (hard) are defined as {\em densities on those  argument variables} where they have simple functional forms and then transformed to \pt\ via appropriate Jacobians as necessary.  The carets indicate unit-normal functions. \ppb\ spectra are plotted here as densities on \pt\ (as published) vs transverse rapidity $y_{t} = \ln[(p_t + m_{t\pi})/m_\pi]$. For convenience below note that pion \yt\ = 2 corresponds to $p_t \approx 0.5$ GeV/c, \yt\ = 2.7 to 1 GeV/c, \yt\ = 4 to 3.8 GeV/c and \yt\ = 5 to 10 GeV/c. The spectrum analysis described here is presented in greater detail in Ref.~\cite{pidsss}.

\begin{figure}[h]
	\includegraphics[width=1.62in]{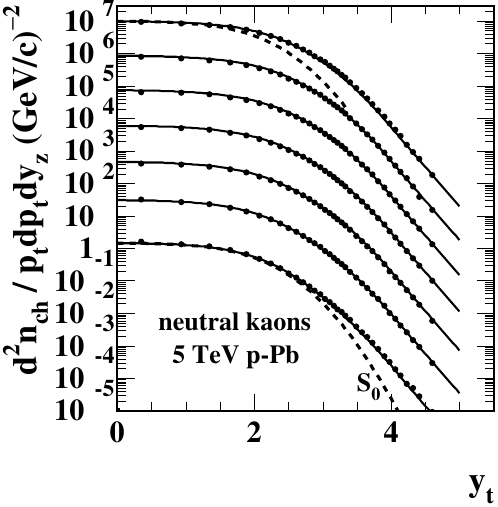}
	\includegraphics[width=1.65in]{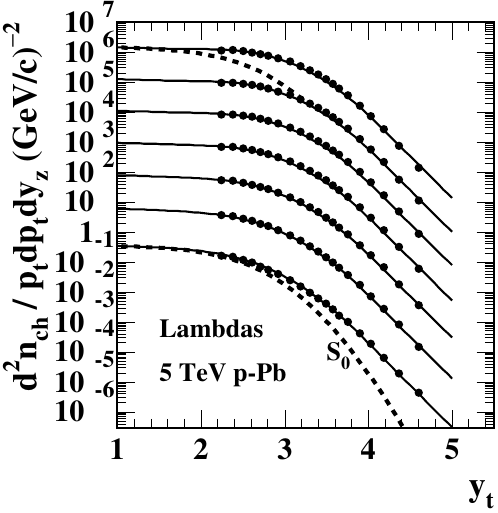}
	\put(-145,105) {\bf (a)}
	\put(-23,105) {\bf (b)}
	\\
	\includegraphics[width=1.65in]{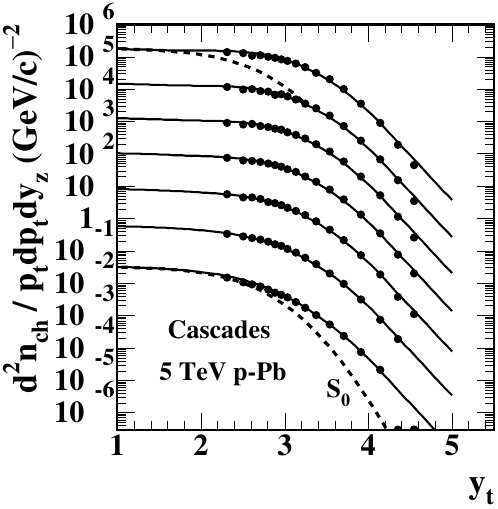}
	\includegraphics[width=1.65in]{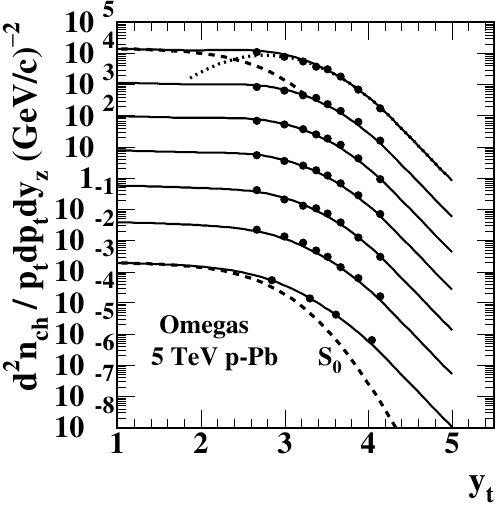}
	\put(-145,105) {\bf (c)}
	\put(-23,105) {\bf (d)}
	\\
	\caption{\label{ppbdata}
		\pt\ spectrum data (points) for strange hadrons from 5 TeV \ppb\ collisions:
		(a) Neutral kaons and
		(b) Lambdas from Ref.~\cite{aliceppbpid},
		(c) Cascades and
		(d) Omegas from Ref.~\cite{alippbss}.
		Solid curves represent the TCM. Dashed curves represent TCM soft components in the form $z_{si}(n_s) \bar \rho_s \hat S_0(p_t)$. The dotted curve in (d) is   the TCM hard component in the form $z_{hi}(n_s) \bar \rho_h \hat H_0(p_t)$.
	}  %  alippbss1aa, ab, ac, ad
\end{figure}

Figure~\ref{ppbdata} shows \pt\ spectra (points) for (a) $K_\text{S}^0$, (b) Lambdas (c) Cascades and (d) Omegas from seven event classes of 5 TeV \ppb\ collisions. The $K_\text{S}^0$ spectra extend down to zero \pt. The solid curves are the full TCM, the dashed curves are fixed soft component $\hat S_0(p_t)$ and the dotted curve is hard component $\hat H_0(p_t)$ for Omegas.

%%%%%%%%%%
%%%%%%%%%%%%

\begin{figure}[h]
	\includegraphics[width=1.65in]{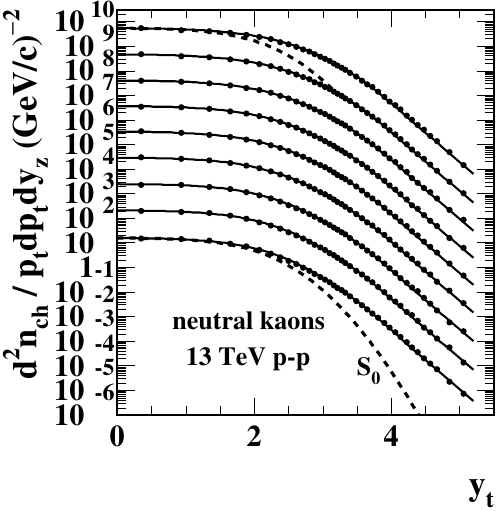}
	\includegraphics[width=1.65in]{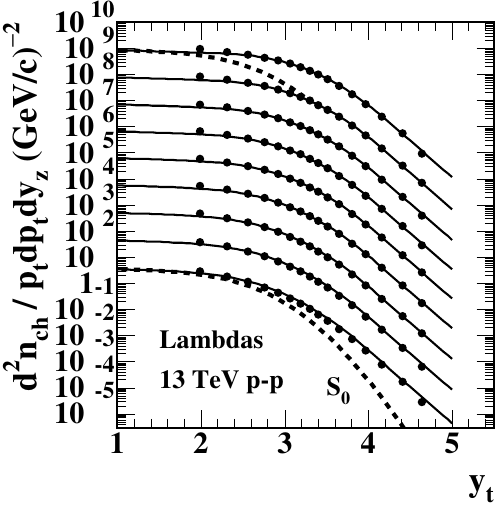}
	\put(-145,105) {\bf (a)}
	\put(-23,105) {\bf (b)}
	\\
	\includegraphics[width=1.65in]{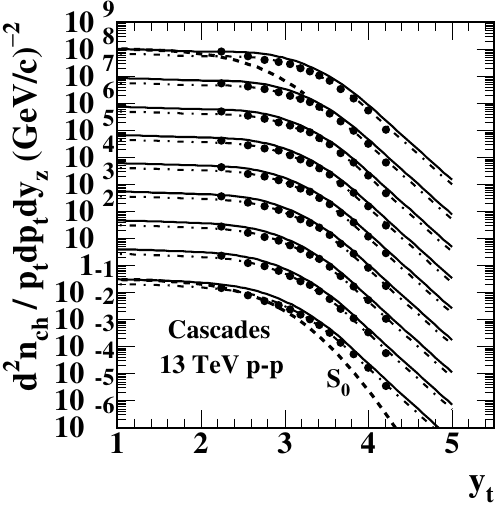}
	\includegraphics[width=1.65in]{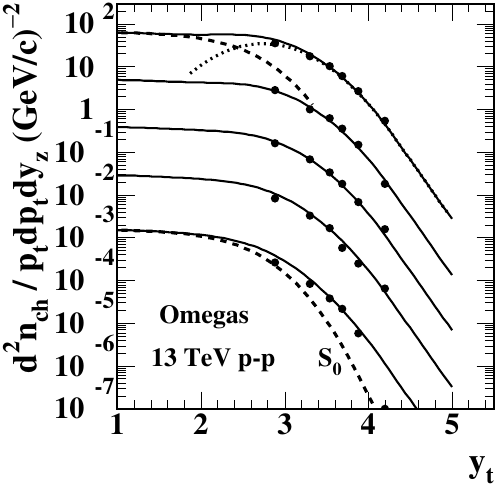}
	\put(-145,105) {\bf (c)}
	\put(-23,105) {\bf (d)}
	\\
	\caption{\label{ppdata}
		\pt\ spectrum data (points) for strange hadrons from 13 TeV \pp\ collisions:
		(a) Neutral kaons,
		(b) Lambdas,
		(c) Cascades and
		(d) Omegas from Ref.~\cite{alicepppid}.
		Solid curves represent the PID spectrum TCM. Dash-dotted curves in (c) are TCM curves reduced by  2/3. Dashed curves are the TCM soft component in the form $z_{si}(n_s) \bar \rho_s \hat S_0(y_t)$.  The dotted curve in (d) is   the TCM hard component in the form $z_{hi}(n_s) \bar \rho_h \hat H_0(p_t)$.
	} %  alippss1aa, ab, ac, ad
\end{figure}

Figure~\ref{ppdata} shows similar results from ten event classes of 13 TeV \pp\ collisions. The published Cascade spectra fall systematically below the TCM prediction whereas published Cascade spectrum integrals are consistent with  the TCM. There is no such discrepancy for \ppb\ Cascades.

%%%%%%%%%%
%%%%%%%%%%%%

The TCM is an example of {\em lossless data compression}: many data values are reduced to a few  parameter values that nevertheless retain all significant information carried by data in a physically-interpretable form. Several particular instances of such compression are noted below.

%%%%%%%%%%%%%%%
\section{$\bf p$-$\bf Pb$ Centrality} \label{cent}

A conventional approach to A-B centrality determination is based on classical Glauber simulations of nuclear collisions. An example is Ref.~\cite{aliceglauber} applied to \ppb\ collisions. An alternative approach combines the TCM with ensemble-mean \mmpt\ data to determine centrality~\cite{tomglauber}.

A TCM for unidentified hadrons is expressed by
\bea
\bar \rho_0(p_t,n_s) = \frac{N_{part}}{2}\bar \rho_{sNN} \hat S_0(p_t) + N_{bin} \bar \rho_{hNN} \hat H_0(p_t)
\eea
that includes $N_{part}$ and $N_{bin}$ as defined above and average densities $\bar \rho_{xNN}$ for individual nucleon pairs. The relation $\bar \rho_{hNN} \approx \alpha \bar \rho_{sNN}^2$ inferred from data~\cite{ppprd} is of fundamental importance. Given that spectrum model the integrated total \pt\ within some angular acceptance is
\bea
\bar P_t =  \frac{N_{part}}{2} n_{sNN} \bar p_{ts} + N_{bin} n_{hNN} \bar p_{th},
\eea
where $\bar p_{tx}$ are determined by model functions $\hat X_0(p_t)$. It is conventional to divide total $\bar P_t$ by total \nch\ within the angular acceptance to calculate $\bar p_t$. However, simplification arises if only the soft component of \nch\ is divided:
\bea \label{meanpt}
\frac{\bar P_t }{n_s} = \bar p_{ts} + x(n_s)\nu(n_s)\bar p_{th},
\eea
where soft multiplicity $n_s = (N_{part}/2)n_{sNN}$ serves also as an event-class index. Given those relations it follows that $N_{part}/2 = \alpha \bar \rho_s / x(n_s)$, $N_{part} = N_{bin} + 1$ and $\nu \equiv 2N_{bin}/N_{part} \leq 2$ completely defines \pa\ centrality.

\begin{figure}[h]
	\includegraphics[width=1.65in]{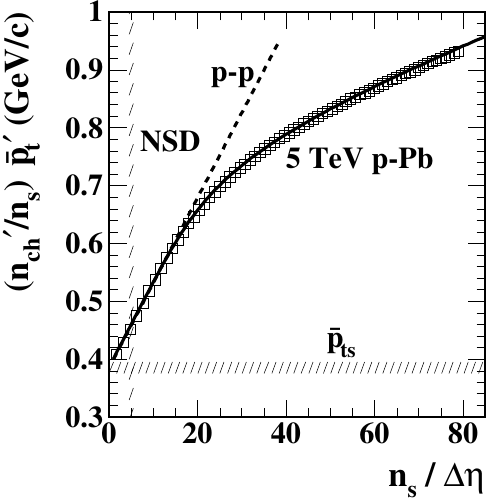}
	\includegraphics[width=1.65in]{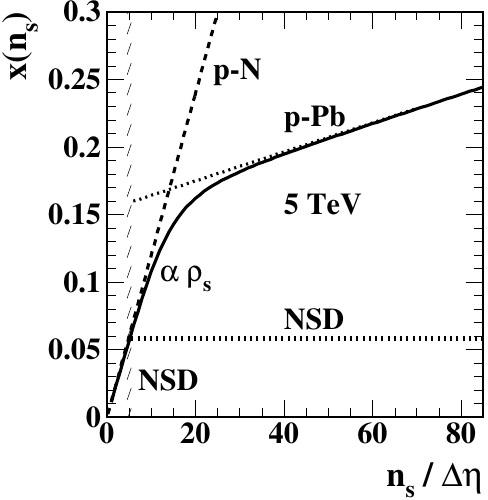}
	\caption{\label{xmpt}
		Left: Ensemble-mean $\bar p_t$ data (points) from 5 TeV \ppb\ collisions as in Ref.~\cite{alicempt} vs soft charge density $\bar \rho_s = n_s / \Delta \eta$. The dashed line represents \pp\ data increasing $\propto \bar \rho_s$ above $\bar p_{ts}$. The solid curve is the TCM described by Eq.~(\ref{meanpt}).
		Right: \nn\ hard/soft ratio $x(n_s)$ corresponding to the TCM curve at left with parameters $\bar \rho_{0s} = 15$ and $m_0 = 0.10$.
	} %  alice3aa4, 1
\end{figure}

Figure~\ref{xmpt} (left) shows ensemble-mean \mmpt\ data for 5 TeV \ppb\ collisions (boxes)~\cite{alicempt}. The dashed line represents corresponding \pp\ data based on Eq.~(\ref{meanpt}) with $\nu = 1$ and $x(n_s) \equiv \bar \rho_{hNN} / \bar \rho_{sNN} \approx  \alpha \bar \rho_{sNN} \rightarrow \alpha \bar \rho_{s}$. The \ppb\ data follow that trend up to a transition point designated by $\bar \rho_{s0}$. Beyond that point \mmpt\ increases with reduced slope, the combination suggesting a complete model for $x(n_s)$.

%%%%%%%%%%
%%%%%%%%%%%%

Figure~\ref{xmpt} (right) shows the $x(n_s)$ model inferred from \mmpt\ data. $x(n_s)$ follows the \pn\ trend $\alpha \bar \rho_s$ up to transition point $\bar \rho_{s0}$ and then continues linearly with reduced slope (factor $m_0$) beyond that point. The solid TCM curve at left corresponds to $\bar \rho_{s0} = 15$ and $m_0 = 0.10$. \mmpt\ data are thus described within point-to-point uncertainties~\cite{tommpt}.

\begin{figure}[h]
	\includegraphics[width=1.65in]{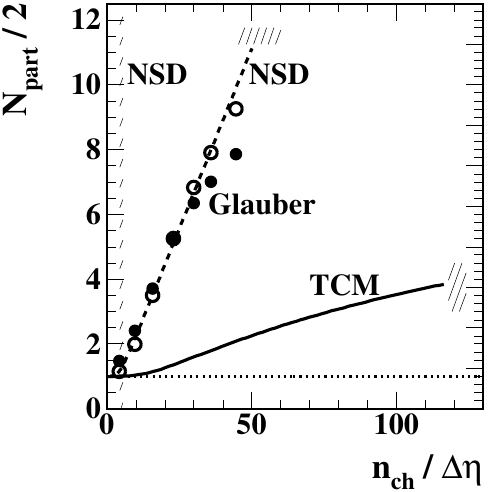}
	\includegraphics[width=1.65in]{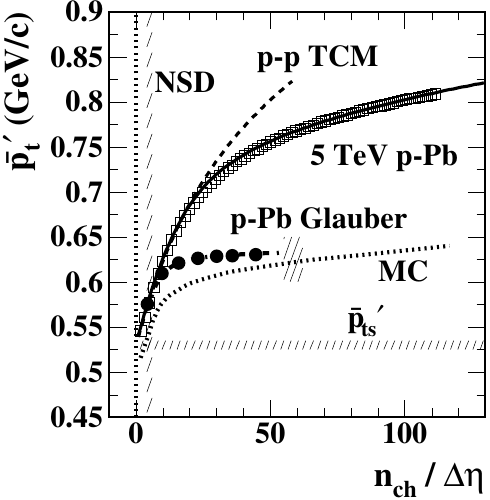}
	\caption{\label{badcent}
		Left: Number of nucleon participant pairs $N_{part}/2$ vs charge density $\bar \rho_0 = n_{ch}/\Delta \eta$ for the TCM (solid) and as inferred in Ref.~\cite{aliceglauber} for the Glauber model (dashed line and solid dots). Hatched bands indicate limits of corresponding trends.
		Right: Results from Fig.~\ref{meanpt} (left) plotted vs $\bar \rho_0$ (solid, dashed, open points) and \mmpt\ values (solid points, dash-dotted curve) predicted by the Glauber trend (solid points at left).
	} %  alicron22g,22gx
\end{figure}

Figure~\ref{badcent} (left) provides a direct comparison between the TCM result and a classical Glauber approach as described in Ref.~\cite{aliceglauber} that assumes $N_{part} \propto n_{ch}$ (dashed line) with published $N_{part}$ values given by the solid points. The TCM analysis in contrast leads to the solid curve consistent with $x(n_s)$ described by the solid curve in Fig.~\ref{xmpt} (right). It is notable that the Glauber description terminates at $\bar \rho_0 \approx 45$ whereas the \mmpt\ data and TCM description extend out to $\bar \rho_0 \approx 115$ (hatched bands).

%%%%%%%%%%
%%%%%%%%%%%%

Figure~\ref{badcent} (right) illustrates the consequences of the two approaches. The assumption $N_{part} \propto n_{ch}$ implies that \nn\ multiplicities $n_{xNN}$ do not vary with centrality. In that case $x(n_s)$ is a fixed quantity in Eq.~(\ref{meanpt}) and only $\nu(n_s)$ may vary. But that quantity has an upper bound of 2 for \pa\ collisions. The trend for conventional $P_t / n_{ch}$ is therefore the solid points in the right panel strongly disagreeing with data (open squares). The source of the large difference has to do with {\em nuclear exclusivity}.

\begin{figure}[h]
	\includegraphics[width=1.55in]{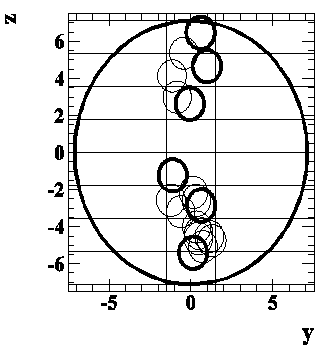}
	\includegraphics[width=1.65in]{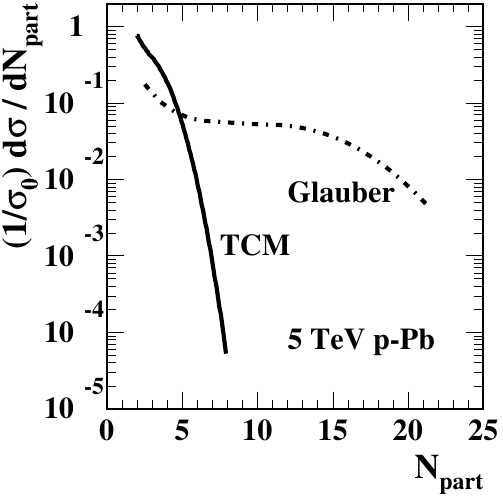}
	\caption{\label{glauber}
		Left: Glauber Monte Carlo event for proton projectile at $b = 0$ incident on Pb nucleus (large circle). Small circles are \pn\ {\em encounters}. Bold circles are encounters that count as inelastic collisions according to an {\em exclusivity} requirement.
		Right: Monte Carlo differential cross sections on participant number $N_{part}$ for TCM (solid) and Glauber (dash-dotted).
	} %  glaubermc, alice400a1
\end{figure}

Figure~\ref{glauber} (left) shows one event from a Glauber Monte Carlo where the large circle is a Pb nucleus and a projectile proton is incident with zero impact parameter. Each small circle is a \pn\ {\em encounter} that might be counted as a collision according to the eikonal approximation. Experimental evidence however suggests that the projectile proton interacts with {\em one nucleon at a time} (bold circles).

%%%%%%%%%%
%%%%%%%%%%%%

Figure~\ref{glauber} (right) shows cross-section distributions on $N_{part}$ from Glauber simulations with (solid) and without (dash-dotted) exclusivity imposed. Those trends in turn correspond to solid and dash-dotted curves in Fig.~\ref{badcent} (right). \mmpt\ data thus {\em require} exclusivity as does the measured trend $\bar \rho_{hNN} \approx \alpha \bar \rho_{sNN}^2$. The result corresponding to the eikonal approximation would be $\bar \rho_{hNN} \approx \alpha \bar \rho_{sNN}^{4/3}$.

The trend for ensemble-mean \mmpt\ data is critical to \ppb\ centrality determination because of the structure of Eq.~(\ref{meanpt}): \mmpt\ data variation above the soft mean $\bar p_{ts}$ is a direct measure of jet production and hard/soft ratio $x(n_s)\nu(n_s)$. The relevant algebra is empirically derived from \pt\ spectra~\cite{ppprd}, not based on physical assumptions.

%%%%%%%%%%%%%%%%%
\section{Nuclear Modification factors}

Nuclear modification factors (NMFs) are rescaled spectrum ratios intended to reveal jet modification that could arise from formation of a dense medium (i.e.\ QGP) in high-energy nuclear collisions. Certain NMF trends might then serve to indicate the presence of a QGP. Before discussing NMFs it is informative to consider spectrum hard components revealed by TCM analysis.

\begin{figure}[h]
	\includegraphics[width=1.65in]{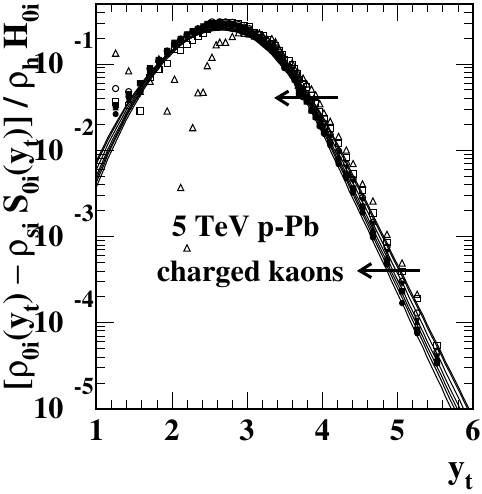}
	\includegraphics[width=1.65in]{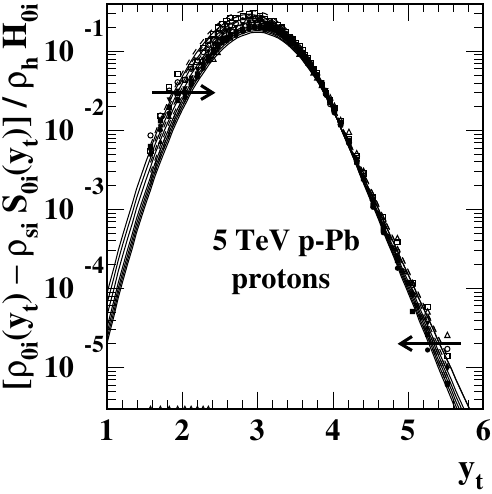}
	\caption{\label{harddat}
		Spectrum hard components for charged kaons (left) and protons (right) from 5 TeV \ppb\ collisions for data (points) and TCM (curves) as reported in Ref.~\cite{ppbnmf}. Arrows indicate TCM model shifts with increasing event centrality.
	}  % alippbhi630aa4st, 630aa4st
\end{figure}

Figure~\ref{harddat} shows spectrum hard components (points) for charged kaons (left) and protons (right) from 5 TeV \ppb\ collisions as reported in Ref.~\cite{ppbnmf}. The curves show the corresponding TCM that describes data within point-to-point uncertainties (with the exception of the most peripheral kaon spectra). Systematic variation with increasing centrality is indicated by the arrows. Note that meson and baryon trends are typically qualitatively different. Hard components presented in this way carry all available information on jet contributions to hadron spectra including any ``jet modifications''

%%%%%%%%%%
%%%%%%%%%%%%

NMFs are conventionally defined as
\bea
R_\text{pPb} = \frac{\bar \rho_\text{0pPb}(p_t,n_s)}{N_{bin} \,\bar \rho_\text{0pp}(p_t,n_s)},
\eea
but given the issue with determination of $N_{bin}$ noted in the previous section interpretation of NMFs so defined may be problematic. It is informative to examine unrescaled spectrum ratios (denoted by $R'_\text{pPb}$) in detail.
\bea \label{rpbp}
R_{p\text{Pb}}' &=& \frac{ z_{si}(n_s) \bar \rho_{s} \hat S_{0i}(p_t) +   z_{hi}(n_s) \bar \rho_{h} \hat H_{0ip\text{Pb}}(p_t,n_s)}{z_{sipp} \bar \rho_{spp} \hat S_{0i}(p_t) + z_{hipp} \bar \rho_{hpp} \hat H_{0ipp}(p_t)}~~
\\ \nonumber 
&\rightarrow& \frac{ z_{si}(n_s) (N_{part}/2) \bar \rho_{sNN} \hat S_{0ipPb}(p_t)}{z_{sipp} \bar \rho_{spp} \hat S_{0ipp}(p_t)} ~~\text{for low \pt}
\\ \nonumber 
&\rightarrow& \frac{z_{hi}(n_s) N_{bin} \bar \rho_{hNN} \hat H_{0ip\text{Pb}}(p_t,n_s)}{z_{hipp} \bar \rho_{hpp} \hat H_{0ipp}(p_t)}~~\text{for high \pt},
\eea

\begin{figure}[h]
	\includegraphics[width=1.65in]{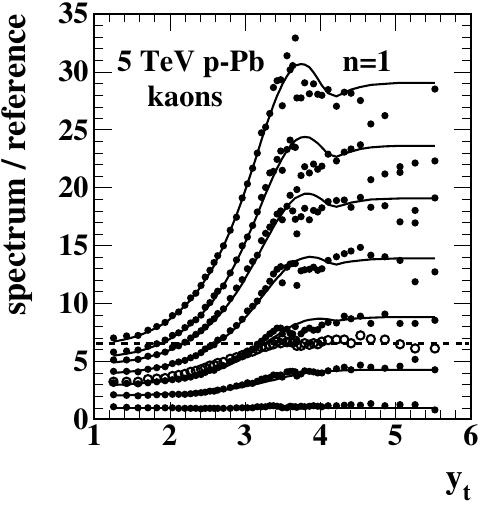}
	\includegraphics[width=1.65in]{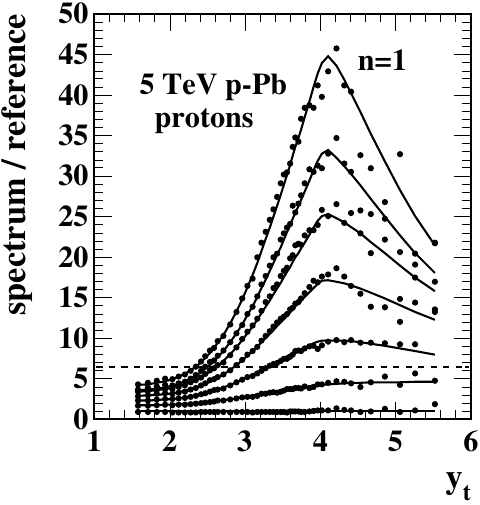}
	\caption{\label{specrat}
	Spectrum ratios $R'_{pPb}$ for charged kaons (left) and protons (right) and for data (points) and TCM (curves) from 5 TeV \ppb\ collisions. The reference spectra (denominators) are TCM peripheral ($n = 7$) spectra.
	}  % alippbhi10bb, cc
\end{figure}

Figure~\ref{specrat} shows spectrum ratios $R'_\text{pPb}$ (points) for kaons (left) and protons (right) from 5 TeV \ppb\ collisions as reported in Ref.~\cite{ppbnmf}. Solid curves are the corresponding TCM. Reference spectra in the denominator of $R'_\text{pPb}$ are in this case the TCM for the most peripheral event class rather than \pp\ spectra. It is of further interest to see how the detailed structure of Eq.~(\ref{rpbp}) affects $R'_\text{pPb}$.

%%%%%%%%%%
%%%%%%%%%%%%

In the low-\pt\ limit model functions $\hat S_{0ix}$ always cancel as a manifestation of nuclear transparency~\cite{transparency}: projectile-nucleon dissociation (and hence $\hat S_{0i}$) is insensitive to the collision process. According to  the assumption $N_{part} \propto n_{ch}$ in Ref.~\cite{aliceglauber} densities $\bar \rho_{sx}$ must be independent of centrality and thus also cancel. The factors $z_{si}$ are $O(1)$ and may be ignored here for the sake of argument. In what follows the hard-component model functions are held fixed and thus also cancel in the high-\pt\ limit. What remain then are $N_{part}/2$ at low \pt\ and $N_{bin}$ at high \pt.

\begin{figure}[h]
	\includegraphics[width=1.65in]{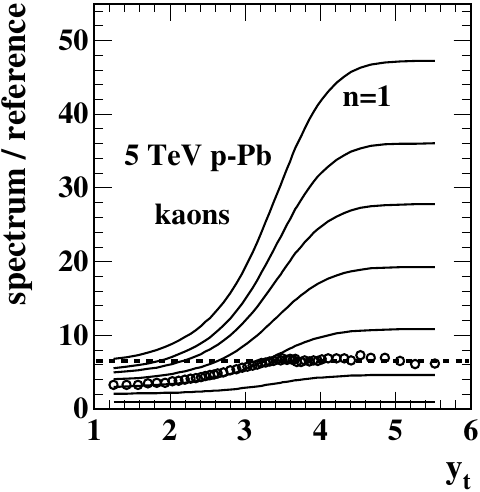}
	\includegraphics[width=1.65in]{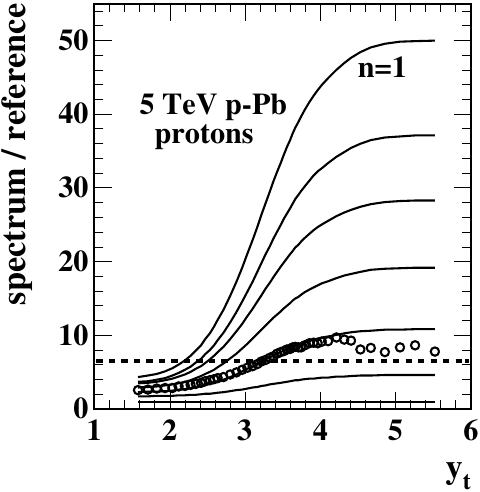}
	\caption{\label{fixedhard}
	TCM $R'_{pPb}$ model curves for charged kaons (left) and protons (right) from 5 TeV \ppb\ collisions. Hard-component models $\hat H_0(p_t)$ in Eq.~\ref{rpbp} are here held fixed independent of centrality. The points are explained in the text.
	} %  alippbhi10bboff, ccoff
\end{figure}

Figure~\ref{fixedhard} shows TCM $R'_\text{pPb}$ curves with hard components $ \hat H_{0i}$ held fixed. The high-\pt\ to low-\pt\ ratio for $n=1$ protons is greater than 10, but the corresponding ratio for  $R'_\text{pPb}$ is $2N_{bin}/N_{part} \equiv \nu \leq 2$. The contradiction arises from the assumption $N_{part} \propto n_{ch}$; in fact \nn\ charge densities vary strongly with \ppb\ centrality,  hard components especially so because of  relation $\bar \rho_{hNN} \approx \alpha \bar \rho_{sNN}^2$.

%%%%%%%%%%
%%%%%%%%%%%%

The open points correspond to the ratio of NSD \ppb\ spectra to minimum-bias \pp\ spectra as reported in Ref.~\cite{alicenucmod}. At lower \pt\ the ratios correspond to \ppb\ event class $n = 5$. The dashed lines correspond to Glauber estimate $N_{bin} \approx 6.5$ (see the third solid point in Fig.~\ref{badcent}, left) which seems to correspond in turn with $R'_\text{pPb}$, for kaons at least, and thus to $R_\text{pPb} \approx 1$ at higher \pt. However, the correct value of $N_{bin}$ for that event class, derived from \mmpt\ data as described in Sec.~\ref{cent}, is 1.3. The numerical agreement in that particular case is thus accidental, and the approach fails dramatically for other event classes.

The only difference between TCM curves in Figs.~\ref{specrat} and \ref{fixedhard} is the hard-component models maintained fixed in the latter case. The source of  the difference is simply determined by the data in Fig.~\ref{harddat}. Small shifts in the widths of hard components coupled to steeply-falling \pt\ dependence lead to large changes in spectrum {\em ratios}  Fig.~\ref{specrat}. The latter figure by itself cannot reveal what the causes are, whereas the differential hard components in Fig.~\ref{harddat} show the small width shifts that {\em are} suitable subjects for further study in a QCD context. Note that kaon ratios are flat above $y_t = 3.5$ because the shifts in Fig.~\ref{harddat} (left) do not vary with centrality, while the sharply-peaked structure for protons arises from the opposing shifts in Fig.~\ref{harddat} (right). The same sharp peak is seen in $p/\pi$ spectrum ratios where it is typically attributed to radial flow. Ratios tend to discard critical information leading to results that are uninterpretable. For full access to data information the {\em factorization} inherent in the TCM is essential.

%%%%%%%%%%%%%%%%%
\section{Strangeness enhancement}

One of the earliest proposed indicators for formation of a QGP in high-energy nuclear collisions was strangeness enhancement~\cite{mullerqgp} expected to follow from quark deconfinement in a hot and dense medium. While strangeness enhancement was reportedly observed in \auau\ collisions at RHIC, within the past ten years strangeness enhancement has been claimed as well for \ppb\ collisions~\cite{alippbss} and even \pp\ collisions~\cite{alicestrange}. In this section the TCM is used to investigate data trends said to support such claims.

Equation~(\ref{pidspectcm}) is the spectrum TCM for identified hadrons, where $\bar \rho_{si} = z_{si}(n_s) \bar \rho_s$ and $\bar \rho_{hi} = z_{hi}(n_s) \bar \rho_h$ include fractional abundances $z_{xi}(n_s) \leq 1$ defined by
\bea \label{zsix}
z_{si}(n_s) &=& z_{0i}(n_s) \frac{1 + x(n_s) \nu(n_s)}{1 + \tilde z_i(n_s)x(n_s) \nu(n_s)}
\\ \nonumber
z_{hi}(n_s) &=& \tilde z_i(n_s)z_{si}(n_s),
\eea
where $z_{0i}$ is the total fractional abundance for hadron species $i$ that may be compared with statistical-model predictions. Coefficients $z_{si}(n_s)$ and $z_{hi}(n_s)$ have been measured for \ppb\ and \pp\ data and define the TCM in Figs.~\ref{ppbdata} and \ref{ppdata} that describes data within statistical uncertainties~\cite{pidsss}.

\begin{figure}[h]
	\includegraphics[width=3.3in]{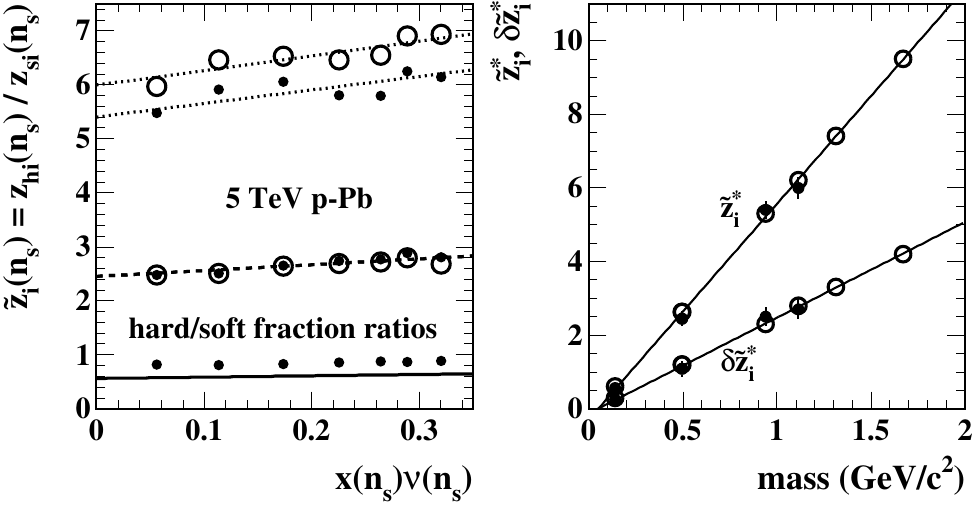}
	\caption{\label{ztilde}
		Left: Ratios $\tilde z_{i}(n_s) = z_{hi}/z_{si}$ inferred from $z_{si}$ and $z_{hi}$ entries reported in Ref.~\cite{pidpart1} for charged (solid dots) and neutral (open circles) hadrons. The lines are linear parametrizations $\tilde z_i = \tilde z_i^* + \delta \tilde z_i^* x(n_s) \nu(n_s)$ that describe the ratio data: solid, dashed and dotted for pions, kaons and baryons respectively.
		Right: Coefficients $\tilde z_i^*$ and $\delta \tilde z_i^*$ for linear descriptions of ratio data in the left panel plotted vs hadron mass. The lines represent proportionality to hadron mass. The solid dots are values inferred in Ref.~\cite{pidpart1}. Open circles provide linear extrapolations on hadron mass for $\Xi$ and $\Omega$ in the present study.
	}  % alice680ey
\end{figure}

Figure~\ref{ztilde} (left) shows hard/soft fraction ratio $\tilde z_i(n_s) = z_{hi}(n_s) / z_{si}(n_s)$ based on the measured fractions. When plotted vs hard/soft ratio $x(n_s)\nu(n_s)$ data trends are consistent with linear variations for lighter hadron species.

%%%%%%%%%%
%%%%%%%%%%%%

Figure~\ref{ztilde} (right) shows slopes and intercepts (solid points) for measured trends at left  plotted vs hadron mass that reveal simple proportionality. The open circles are the adopted values for the TCM shown in  Figs.~\ref{ppbdata} and \ref{ppdata}. This is one example of lossless data compression in that 28 parameter values at left are reduced to two numbers at right that predict Cascade and Omega spectra solely based on hadron mass.  Note that Eq.~(\ref{zsix}) (first line) extrapolates to $z_{0i}$ in the limit $x(n_s)\nu(n_s) \rightarrow 0$.

\begin{figure}[h]
	\includegraphics[width=3.3in]{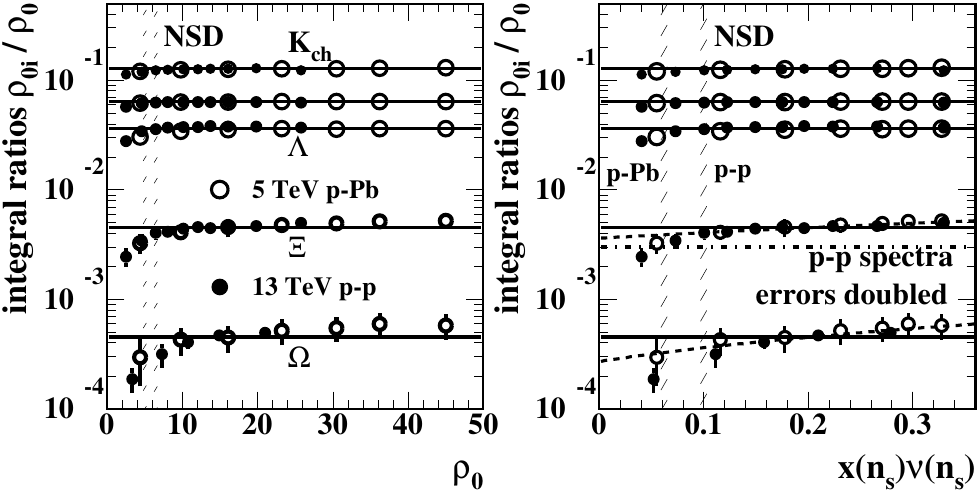}
	\caption{\label{z0trends}
		Left: PID yield ratios $z_{0i}(n_s) \equiv \bar \rho_{0i} / \bar \rho_0$ vs $\bar \rho_0$ based on 13 TeV \pp\ data in Fig.~8 of Ref.~\cite{alippss} (solid dots) and 5 TeV \ppb\ data in Table~4 of Ref.~\cite{alippbss} (open circles) which then represent variable quantity  $z_{0i}(n_s)$ in the TCM context. The \ppb\ Omega errors have been doubled to make them visible. The solid lines are fixed values $z_{0i}^*$ as reported in the present study. The hadron species, from the top, are charged kaons, neutral kaons, Lambdas, Cascades and Omegas.
		Right: Data in the left panel plotted vs TCM hard/soft ratio $x(n_s) \nu(n_s)$ demonstrating statistical equivalence of \pp\ and \ppb\ collision systems for quantity $z_{0i}(n_s)$ plotted vs $x\nu$.
	}  % alippss4a
\end{figure}

Figure~\ref{z0trends} (left) shows total fractional abundances $z_{0i} = \bar \rho_{0i} / \bar \rho_0$ (points) adapted from yield data reported in Refs.~\cite{alippbss} and \cite{alippss}. The bold lines are values obtained from the present study as described above. The plot format $z_{0i}(n_s)$ vs $\bar \rho_0$ is as preferred in the cited references wherein \pp\ and \ppb\ trends seem quite different.

%%%%%%%%%%
%%%%%%%%%%%%

Figure~\ref{z0trends} (right) shows  $z_{0i}(n_s)$ plotted vs hard/soft ratio $x(n_s)\nu(n_s)$ in which format the \pp\ and \ppb\ data appear statistically equivalent. That is another example of data compression in that two collision systems may be described by the same model. In this format ``strangeness enhancement'' appears as the significant positive slope (dashed lines) for Cascades and Omegas. However, relative to published systematic errors (doubled in the plot for visibility) the effect appears at the two-sigma level. Assuming a conventional interpretation for those data the evidence for QGP formation based on strangeness enhancement is nevertheless not compelling.

\begin{figure}[h]
	\includegraphics[width=3.3in]{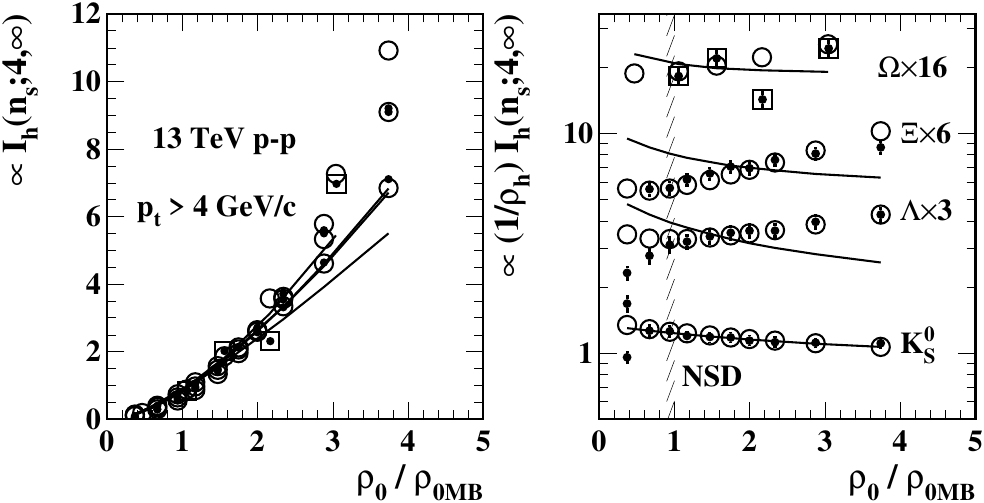}
	\caption{\label{fig6a}
		Left: High-\pt\ PID spectrum integrals ($p_t > 4$ GeV/c) from 13 TeV \pp\ collisions (solid dots) for four hadron species as presented in Fig.~6 (top row) of Ref.~\cite{alippss}. Open squares identify $\Omega$ data. Open circles and curves are TCM results.
		Right: Data and curves in the left panel rescaled by $\bar \rho_h = x(n_s) \bar \rho_s$. Solid dots are data from Ref.~\cite{alippss}. Open circles are derived from TCM spectra (curves) in Fig.~\ref{ppdata}. Solid curves are TCM trends with no hard-component shifts on \yt.
	} % alippss3a
\end{figure}

Figure~\ref{fig6a} (left) shows data from Fig.~6 of Ref.~\cite{alippss} (solid points) wherein integrals of \pt\ spectra above 4 GeV/c are shown for 13 TeV \pp\ collisions. The data are said to be ``self-normalized'' resulting in data clustering as closely as possible. The open circles are from the same procedure applied to TCM spectra, and the solid curves are explained below. That figure contains much more information than is immediately apparent.

The integrals can be described in a TCM context by
\bea \label{pint}
I_{hi}(4 \text{ GeV/c},\infty) \propto z_{hi}(n_s) \bar \rho_h \int_\text{4 GeV/c}^\infty \hspace{-.16in} p_t dp_t \hat H_{0i}(p_t,n_s).
\eea
Factor  $ \bar \rho_h$ ($\sim$ quadratic trend vs $\bar \rho_0$) is extraneous and can be immediately removed with good effect.

%%%%%%%%%%
%%%%%%%%%%%%

Figure~\ref{fig6a} (right) shows the contents of the left panel divided by $ \bar \rho_h \approx \alpha \bar \rho_s^2$. It is now apparent that the TCM (open circles) provides a reasonable description of the data integrals (solid points). If the hard-component model functions are held fixed the resulting solid curves  are proportional to factor $z_{hi}(n_s)$ in Eq.~(\ref{pint})
\bea \label{curly}
z_{hi}(n_s) = z_{0i}(n_s) \left \{ \frac{\tilde z_i(n_s)[1+x(n_s)\nu(n_s)]}{1+\tilde z_i(n_s)x(n_s)\nu(n_s)}\right\}.
\eea

\begin{figure}[h]
	\includegraphics[width=1.65in]{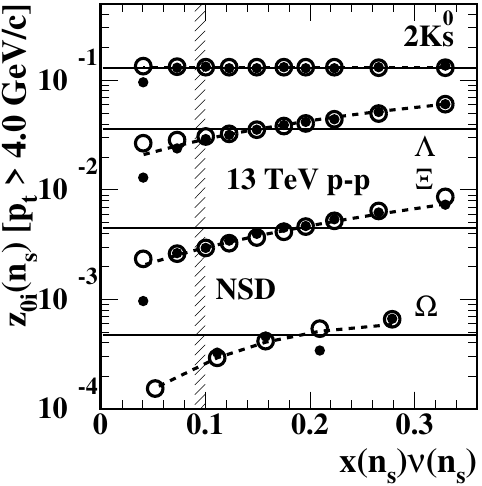}
\includegraphics[width=1.62in]{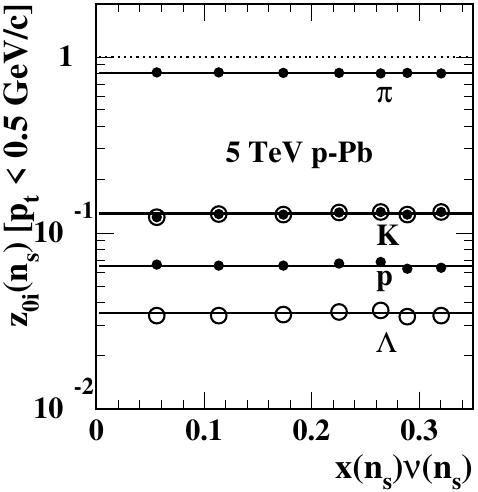}
	\caption{\label{fig6b}
		Left: Rescaled integrals $I_h / \bar \rho_h$ of Fig.~\ref{fig6a} (right) transformed according to the curly brackets in Eq.~(\ref{curly}).
		Right: A similar procedure applied to spectrum integrals over \pt\ interval [0,0.5] GeV/c showing no significant $z_{0i}$ variation.
	} %  alippss3c, alice680cc
\end{figure}

Figure~\ref{fig6b} (left) shows the contents of Fig.~\ref{fig6a} (right) divided by the expression in curly brackets above. The results are then simply proportional to the hard-component integrals in Eq.~(\ref{pint}). The result for each hadron species has been rescaled to $z_{0i}$ values corresponding to that species. The issue here is the {\em relative} variation on $x\nu$. There is no significant variation for mesons $K^0_\text{S}$ but {\em equal} substantial variations for three baryon species. In particular, there is no apparent dependence on strangeness {\em per se}. The next question is how these results at higher \pt\ relate to the full \pt\ integrals in Fig.~\ref{z0trends} (right).

%%%%%%%%%%
%%%%%%%%%%%%

Figure~\ref{fig6b} (right) shows trends at low \pt\ indicating no significant variation, presumably because there is no jet contribution there. The variations in Fig.~\ref{z0trends} (right) then result from a linear combination of no change at lower \pt\ and Fig.~\ref{fig6b} (left) at higher \pt\ (a jet contribution). The result then depends on how ``jetty'' a given hadron species is, but that is determined by $\tilde z_i(n_s)$. From Fig.~\ref{ztilde} (right) $\tilde z_i(n_s)$ is 0.6 for pions and 10 for Omegas. Close examination of Fig.~\ref{ppdata} (d) shows that for most-central data every Omega detected is a jet fragment (lies on the dotted hard-component curve). The variations in Fig.~\ref{z0trends} (right) (for baryons only) are then the trends in Fig.~\ref{fig6b} (left) scaled down by $\tilde z_i(n_s)$: the $z_{0i}(n_s)$ variations are a jet effect having no apparent correlation with strangeness.

%%%%%%%%%%%%
\section{CMS ridge}

\begin{figure}[h]
	\includegraphics[width=3.3in]{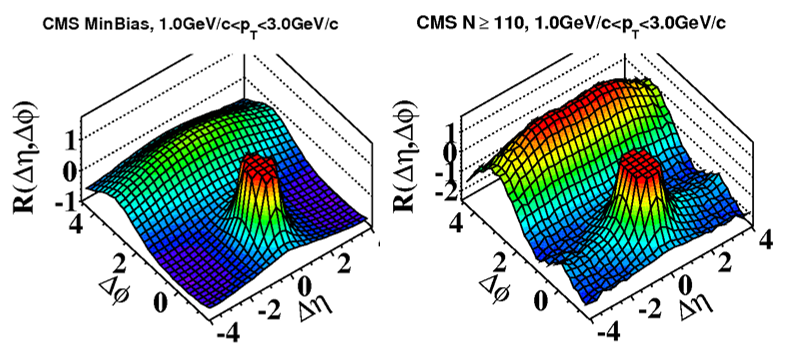}
	\caption{\label{cmsridge}
	2D angular correlations for low \nch\ (left) and high \nch\ (right) from 13 TeV \pp\ collisions as reported in Ref.~\cite{cmsridge}.
	}  % cms-ridge.png
\end{figure}

Figure~\ref{cmsridge} shows 2D angular correlations for 13 TeV \pp\ collisions reported by the CMS collaboration~\cite{cmsridge} for low (left) and high (right) event multiplicity and with a \pt\ cut applied as noted on the panels. The structure of note is the ``ridge'' apparent at the azimuth origin on the two sides of the jet peak at right. That structure has been interpreted as representing long-range correlations associated with ``collectivity'' (flow), in this case apparently in \pp\ collisions. As such it deserves close attention. A follow-up study of 200 GeV \pp\ collisions was reported in Ref.~\cite{ppquad} wherein 2D angular correlations were studied within a TCM context. The results are described below.

%%%%%%%%%%
%%%%%%%%%%%%

\begin{figure}[h]
	\includegraphics[width=1.65in]{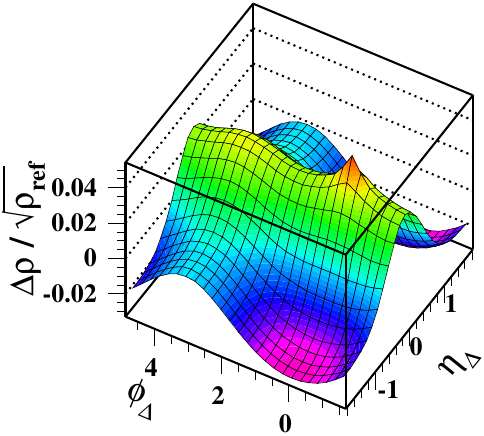}
	\includegraphics[width=1.65in]{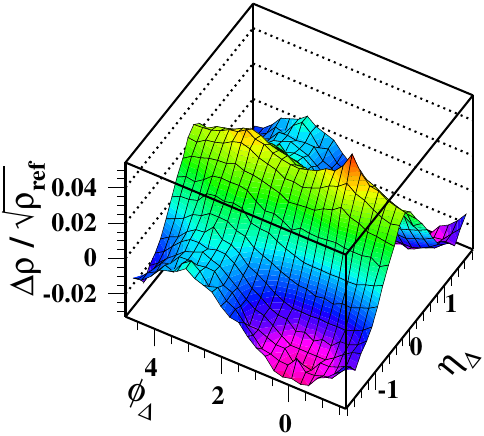}
	\put(-145,95) {\bf (a)}
	\put(-23,95) {\bf (b)}
	\\
	\includegraphics[width=1.65in]{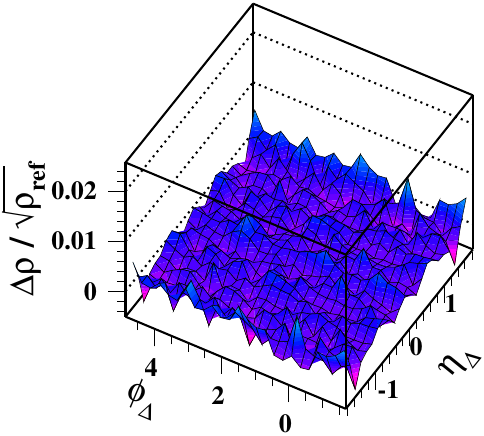}
	\includegraphics[width=1.65in]{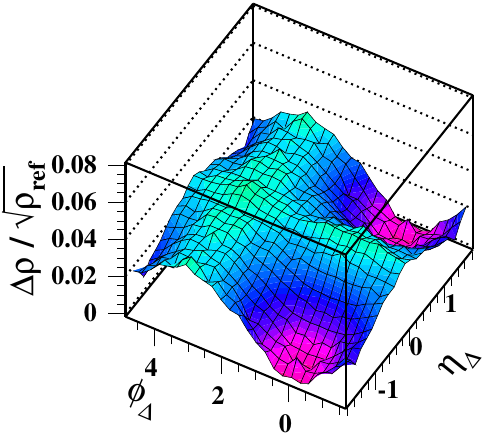}
	\put(-145,95) {\bf (c)}
	\put(-23,95) {\bf (d)}
	\caption{\label{lownch}
Perspective views of \yt-integral 2D angular correlations as $\Delta\rho/\sqrt{\rho_{ref}}$ on  $(\eta_{\Delta},\phi_{\Delta})$  from  \pp\ collisions at $\sqrt{s}$ = 200~GeV for low event multiplicity reported in Ref.~\cite{ppquad}. (a) fit model, (b) data histogram, (c) fit residuals (vertical sensitivity increased two-fold), (d) jet + NJ quadrupole contributions obtained by subtracting fitted offset, soft-component and BEC/electron model elements from the data histograms. 
	} % ppcms23-0ax, bx, cx, dx
\end{figure}

Figure~\ref{lownch} shows  2D angular correlations for a low-\nch\ event class of 200 GeV \pp\ collisions including  (a) TCM fit model, (b) data, (c) fit residuals and (d) data minus soft and Bose-Einstein (BEC) models. The quality of the fit is indicated by the fit residuals (c) consistent with Poisson statistics. The soft component is the broad peak on $\eta$ in (a,b) nearly independent of azimuth. Other model elements include a same-side (SS) 2D jet peak, an away-side (AS) cylindrical dipole (back-to-back jet pairs) and a cylindrical quadrupole. A Bose-Einstein model component is described below.

%%%%%%%%%%
%%%%%%%%%%%%

\begin{figure}[h]
	\includegraphics[width=1.65in]{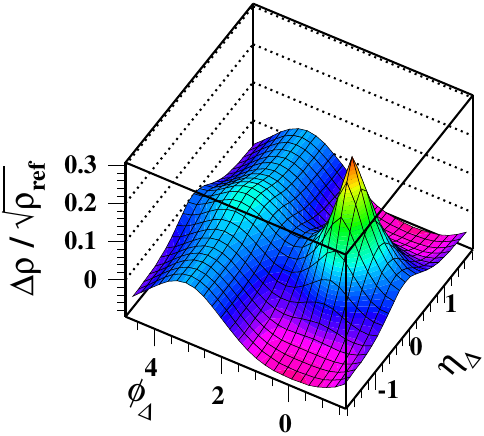}
	\includegraphics[width=1.65in]{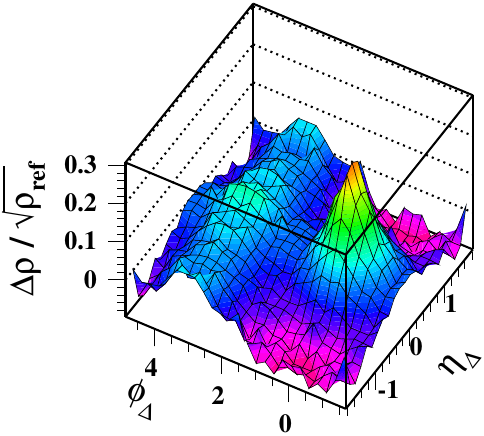}
	\put(-145,95) {\bf (a)}
	\put(-23,95) {\bf (b)}
	\\
	\includegraphics[width=1.65in]{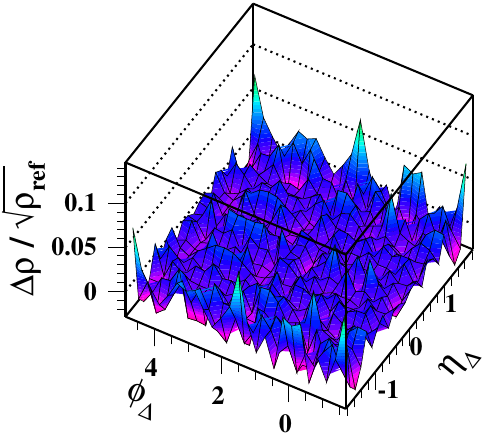}
	\includegraphics[width=1.65in]{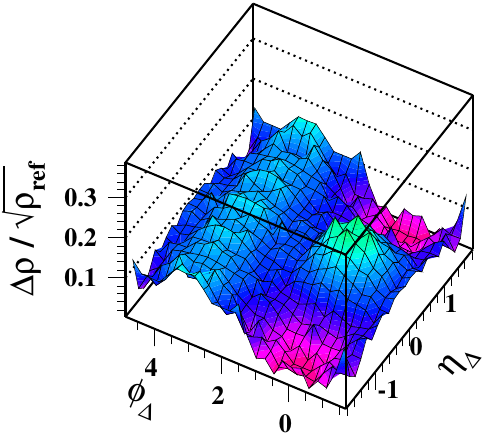}
	\put(-145,95) {\bf (c)}
	\put(-23,95) {\bf (d)}
	\caption{\label{highnch}
	Same as Fig.~\ref{lownch} except for high event multiplicity.
	}  % alicron22g,22gx
\end{figure}

Figure~\ref{highnch} shows  2D angular correlations for a high-\nch\ event class of 200 GeV \pp\ collisions. The panels are as described for Fig.~\ref{lownch}. The Bose-Einstein component is the small 2D exponential atop the broader same-side 2D jet peak ($\approx$ Gaussian) in panel (b). Note that the apparent narrowing on azimuth of the SS 2D jet peak in panel (d) could be misinterpreted. The jet peak there has the same shape as that in Fig.~\ref{lownch} (d). The apparent narrowing is due to a lobe of the quadrupole superposed on the jet peak, a companion manifestation to reduction of positive curvature (``ridge'') at the azimuth origin away from the SS 2D jet peak on $\eta$. Also note the substantial increased negative curvature on  the AS dipole at $\pi$.

The presence or absence of a ``ridge'' in Fig.~\ref{cmsridge} may be discussed in terms of curvatures on azimuth. The curvature of the AS dipole is negative (maximum) at $\pi$ and positive (minimum) at zero. In contrast, the curvature of a cylindrical quadrupole is negative (maxima) at both zero and $\pi$. Thus, if a quadrupole component increases with event \nch\ faster than the AS dipole the AS negative curvature will increase in {\em magnitude} while the positive SS curvature decreases as in the right panel. If those trends continue far enough a SS ``ridge'' (negative curvature, maximum) may result as in Fig.~\ref{cmsridge} (right) (but note the much-increased negative curvature of the AS dipole at $\pi$). Evolution from low \nch\ to high \nch\ of the CMS data is thus indicative of a two-lobed quadrupole structure, not a single-lobed ``ridge.'' Qualitative discussion in terms of curvatures cannot replace quantitative analysis of correlation structure in the context of the TCM.

%%%%%%%%%%
%%%%%%%%%%%%

\begin{figure}[h]
	\includegraphics[width=1.65in]{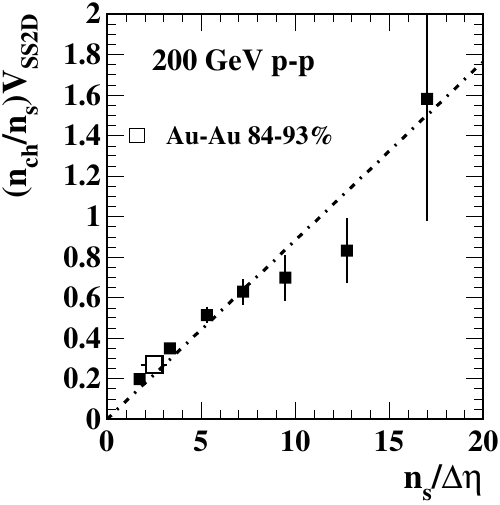}
	\includegraphics[width=1.65in]{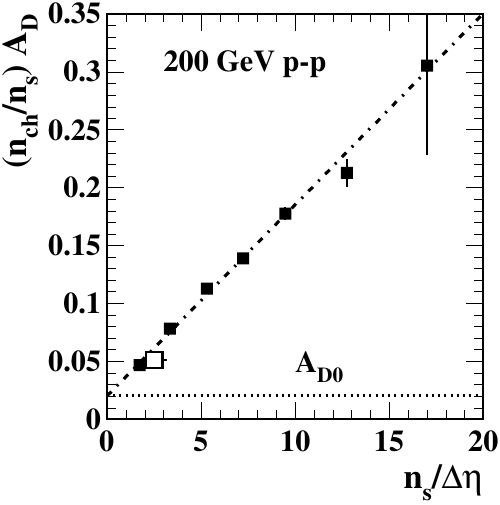}
	\put(-145,35) {\bf (a)}
	\put(-23,35) {\bf (b)}
	\\
	\includegraphics[width=1.65in]{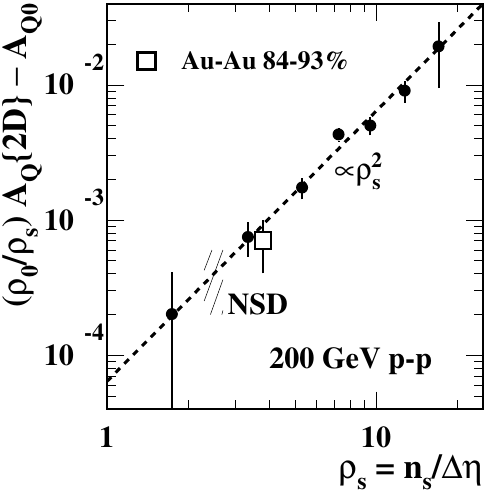}
	\includegraphics[width=1.6in]{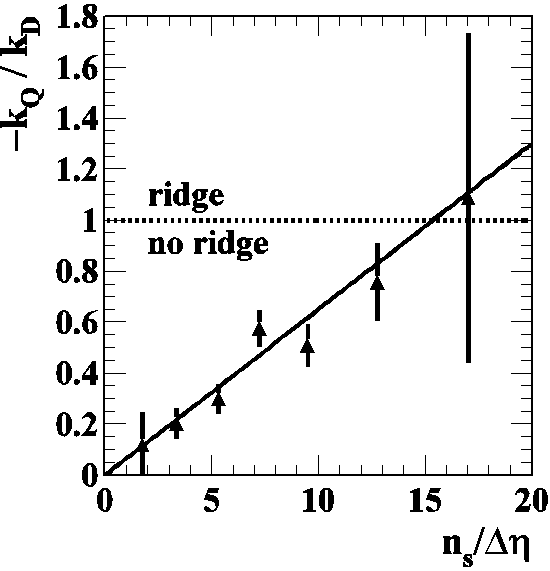}
	\put(-145,40) {\bf (c)}
	\put(-23,35) {\bf (d)}
	\caption{\label{ppcorr}
	(a) SS 2D peak volume $V_\text{SS2D}$ vs soft density $\bar \rho_s = n_{s} / \Delta \eta$. 
	(b)  AS dipole amplitude $A_\text{D}$ vs $\bar \rho_s $
	(c)  NJ quadrupole amplitude $ A_\text{Q} \equiv \bar \rho_s\, v_2^2\{2D\}$ vs $\bar \rho_s$. 
	(d) Quadrupole/dipole curvature ratio $-k_\text{Q} / k_\text{D}$  vs $\bar \rho_s$ and the criterion for a ``ridge.''
	} %  alicron22g,22gx
\end{figure}

Figure~\ref{ppcorr} (a) shows the volume $V_\text{SS2D}$ of the SS 2D jet peak measured as number of correlated pairs scaled by the number of soft hadrons $n_s$ plotted vs $\bar \rho_s = n_s / \Delta \eta$. The linear trend indicates that the peak volume (number of pairs) is $\propto \bar \rho_s^2$. Figure~\ref{ppcorr} (b) shows that the same holds true for the amplitude $A_\text{D}$ of the AS dipole.

%%%%%%%%%%
%%%%%%%%%%%%

Figure~\ref{ppcorr} (c) shows the amplitude $A_\text{Q}\{\text{2D}\}$ (i.e.\ amplitude determined by 2D model fits) of the cylindrical quadrupole plotted on a log-log format and indicating that the quadrupole amplitude (pairs) varies $\propto \bar \rho_s^3$. Note that the quadrupole amplitude {\em increases over three orders of magnitude}, and the $\bar \rho_s^3$ trend extends down below the non-single-diffractive (NSD) event class. Previous measurements for peripheral 200 \auau\ collisions (open boxes) agree with the more-recent \pp\ data.

What physical interpretations can be derived from these {\em quantitative} measurements? First, jet correlations $\propto \bar \rho_s^2$ as in Fig.~\ref{ppcorr} (a) and (b) is simply consistent with the relation  $\bar \rho_h \propto \bar \rho_s^2$ derived from \pt\ spectra as in Sec.~\ref{cent}. But $\bar \rho_s$ represents hadrons produced by projectile nucleon dissociation along the beam axis that may in turn be attributed to low-$x$ gluons as their parents.  $\bar \rho_h \propto \bar \rho_s^2$ may then be interpreted to indicate that dijets result from two-gluon interactions, which should not be surprising for minimum-bias jets near midrapidity. Figure~\ref{ppcorr} (c) tells us that the cylindrical quadrupole is $\propto \bar \rho_s^3$. An analogous argument then suggests that the quadrupole arises from a three-gluon interaction and should be studied as a new form of elementary QCD interaction. Figure~\ref{ppcorr} (d) shows the trend of quadrupole vs dipole curvatures at the azimuth origin and demonstrates quantitatively how the two trends on $n_s$ may lead to a SS ``ridge.''

%%%%%%%%%%%
\section{Summary}\label{summ}

This presentation introduces the following responses to claims of QGP formation  in small collision systems:

With regard to centrality determination for \pa\ collisions, a two-component model (TCM) of hadron production includes factorization of particle densities and \pt\ dependence for individual components that make relevant details quantitatively accessible. Measured values of ensemble-mean \mmpt\ are essential for correct centrality determination: the \mmpt\ data in effect determine the contribution from minimum-bias jets. One result of the method is identification of a novel phenomenon -- {\em exclusivity} for individual \nn\ collisions: only one \nn\ collision {\em at a time}, already implicit from previous \pp\ spectrum analysis. The correspondence with jet production is confirmed by two-particle correlation analysis (see comments on ``the ridge'' below). Incorrect \pa\ centrality arising from classical Glauber analysis causes confusion regarding nuclear modification factors. Comparison of \mmpt\ data trends for \pp, \pa\ and \aa\ collisions is counterintuitive unless analyzed in terms of jet production and exclusivity.

Nuclear modification factors (NMFs) are expected to  reveal the presence of QGP via jet response to a dense medium, but the basic definition is problematic. Rescaling of spectrum ratios by estimated binary-collision number $N_{bin}$ is vulnerable to errors in \pa\ centrality determination as noted above. Spectrum ratios in turn lead to discard of essential information making proper interpretation difficult. Related assumptions such as fixed \nn\ particle densities lead to further difficulties. A limiting case of \pa\ is \pp\ where $N_{part}/2 = N_{bin} = 1$, yet spectrum ratios may change dramatically with increasing event \nch. What should be objects of study are individual spectrum hard components accessible via the TCM.

Strangeness enhancement as a signal of QGP formation emerged within a context of thermalization, fireballs and freezeout -- hadrochemistry as conceptualized at lower collision energies. That narrative lacks an essential element of high-energy nuclear collisions: minimum-bias jet production. The conventional data presentations associated with strangeness enhancement discard critical information as a matter of confirmation bias. Differential analysis via the TCM of data presented in conventional formats actually reveals strong jet contributions that are uncorrelated with strangeness {\em per se} but explain trends attributed to its enhancement. Similar misinterpretations of jet phenomena appearing in spectra and angular correlations can be associated with inferences of radial flow and some aspects of $v_2$ measurements.

A certain feature of \pp\ angular correlations from the LHC has been described as ``the ridge,'' i.e.\ long-range (on $\eta$) correlations near zero azimuth interpreted to indicate ``collectivity'' translated in turn as ``flow.'' Treatment of data is superficial, based on apparent curvatures in limited regions of the angular acceptance. A more complete analysis involving fits to 2D angular correlations over the entire acceptance with an established fit model over a range of event multiplicities reveals critical details. As well as the ``ridge'' (negative curvature) at zero azimuth (same-side) there is greatly increased negative curvature near $\pi$ (away-side). The combination is actually a cylindrical quadrupole, not a single ``ridge.'' The model fits confirm that jet-related correlation structure increases in amplitude as the square of the particle-density soft component, consistent with previous TCM analysis of \pp\ \pt\ spectra, but that the {\em amplitude} of the quadrupole element (number of pairs) increases as the {\em cube} of the soft density {\em over three orders of magnitude}. Given that minimum-bias jets are dominated by a two-gluon interaction these results suggest that the quadrupole arises from a three-gluon interaction. Instead of a signal of collective flow the quadrupole structure actually represents a novel elementary QCD phenomenon.

These four examples demonstrate that analysis methods which discard a substantial fraction of information carried by particle data combined with unjustified assumptions about collision dynamics may lead to confusion susceptible to confirmation bias. There is currently no significant evidence to support claims of QGP formation in small systems and substantial evidence against.

%%%%%%%%%%%%%%%%%%%%%%%%%%%%

\end{document}